# Flower-like $WO_3$-Modified Vulcan Carbon GDEs in Photoelectro-Fenton Process: Efficient Ciprofloxacin Degradation and Mechanistic Insights

*João Paulo C. Moura[1], Vanessa S. Antonin[1], Caio Machado Fernandes[1], Aline B. Trench[1], Erica S. Conrado[2], Michel O. Almeida[3], Kathia M. Honorio[2], Renata Colombo[2], Rafael Sotana[1], Ana M. P. Neto[1], Mauro C. Santos[1*].*

*(1) Laboratório de Eletroquímica e Materiais Nanoestruturados, Centro de Ciências Naturais e Humanas, Universidade Federal do ABC, CEP 09210-170, Rua Santa Adélia 166, Bairro Bangu, Santo André, SP, Brazil*

*(2) University of São Paulo. Av. Ermelino Matarazzo, 1000, São Paulo-SP, Brasil.*

*(3) University of São Paulo. Av. Professor Lineu Prestes, 580, São Paulo-SP, Brasil.*

**Corresponding Author:* Mauro C. Santos

e-mail: mauro.santos@ufabc.edu.br or drmcsa@gmail.com

## Abstract

Herein, the electrochemical synthesis of $H_2O_2$ was investigated using gas diffusion electrodes (GDEs) based on 3% $WO_3$/C electrocatalyst. The 3% $WO_3$/C GDE exhibited superior $H_2O_2$ production, achieving concentrations of 423 mg $L^{-1}$, 586 mg $L^{-1}$, and 916 mg $L^{-1}$ of $H_2O_2$ at 50, 75, and 100 mA $cm^{-2}$, respectively, with an efficiency of ~70% across all applied currents, and a solid improvement in efficiency compared to bare Vulcan carbon. Furthermore, its energy consumption was significantly reduced, attributed to the synergistic interaction between $WO_3$ and Vulcan carbon. The catalytic performance of $WO_3$/C GDE was also assessed in electro-Fenton (EF) processes for ciprofloxacin (CIP) degradation. The EF process demonstrated rapid initial degradation (~70% in 30 min), but a subsequent decline in efficiency due to constraints on iron (II) regeneration. The UV-assisted photoelectro-Fenton process overcame the limitation of sustaining higher degradation rates, achieving complete CIP removal within 90 min. The total organic carbon mineralization was further optimized using boron-doped diamond anodes, achieving up to 66% mineralization. A mechanistic degradation pathway was proposed, highlighting two primary routes: hydroxyl radical attack on the piperazine ring, with or without defluorination, and oxidation of the quinolone ring. Theoretical analysis of byproducts reveals that electrocatalytic oxidation treatment reduces the environmental impact of the resulting wastewater and corroborates the experimental degradation pathways. The results confirm the high efficiency of $WO_3$/C GDEs for both $H_2O_2$ electrogeneration and persistent pollutant degradation, offering a promising approach for environmental remediation.

## 1. Introduction

Fluoroquinolone antibiotics are among the most significant antibacterial agents in human and veterinary treatments. Ciprofloxacin (CIP), a member of the fluoroquinolone family, is a broad-spectrum antibiotic effective against both gram-positive and gram-negative bacteria, widely used to treat infections of the urinary, digestive, and respiratory systems [1]. Despite the healthcare benefits of antibiotics, the overuse and misuse of antimicrobial drugs contribute to the emergence of antimicrobial resistance (AMR). This issue has attracted increasing concern due to its potential threat to human, animal, and environmental health. According to the World Health Organization (WHO), antimicrobial resistance has emerged as one of the top 10 global health threats [2,3]. Combating antimicrobial resistance requires a multifaceted and coordinated approach. Wastewater treatment is critical for addressing AMR, as it helps mitigate the spread of resistant bacteria and antibiotic residues in the environment.

Several physical, biological, and chemical treatment technologies, such as adsorption, activated sludge process, aerobic, and anaerobic bio-digestion [4]and advanced oxidation processes (AOPs) are commonly used. Advanced oxidation processes involve the degradation of organic matter by highly reactive oxidizing species (ROS) generated in the medium, typically via chemical reactions. This approach has been well reported and applied for this purpose [5]. Recently, combining electrochemistry with AOP has gained attention. These electrochemical advanced oxidation processes (EAOPs) reduce or eliminate the need for external chemical additives to degrade persistent organic pollutants, thereby enhancing energy efficiency and degradation capacity. As a result, EAOPs are becoming an increasingly environmentally friendly solution.

The Electro-Fenton (EF) process has proven to be an efficient method for degrading persistent organic pollutants, including dyes [6,7], pesticides [8,9], pharmaceuticals [10–12],[10–12]and synthetic organic chemicals [13,14]. Reported results demonstrate the effectiveness of the EF process, which relies on the *in-situ* electrogeneration of hydrogen peroxide (H2O2) and its activation with iron ions. As a result, significant efforts have been made to develop optimal electrocatalysts capable of efficiently generating H2O2 while consuming less energy. Carbonaceous materials have attracted attention due to their desirable features, such as high electrical conductivity and selective active sites for the 2-electron oxygen reduction reaction (ORR-2e-) pathway [15,16]. Although carbon-based materials are inherently well-suited for facilitating the ORR-2e-, they are not considered optimal electrocatalysts.

Modifying carbon materials with metallic oxides has been reported as an optimal strategy for optimizing H2O2 electrosynthesis. In recent works, Nb2O5[17–19], CeO2[10,20–22], $MnO_2$[23–25] and $WO_3$[7,26,27] have been investigated, demonstrating significant improvement in the ORR-2e- through oxide-carbon coupling. Valim *et al.* [17] used a GDE based on $Nb_2O_5$/C. They achieved a 2-fold increase in $H_2O_2$ accumulation compared to pure carbon black and efficient levofloxacin degradation in solution via the electro-Fenton process. Similarly, Paz *et al.* [6] investigated $WO_{2.72}$ nanoparticles supported by Vulcan XC72 and Printex 6L, substantially improving the ORR-2e- selectivity and applications in organic dye degradation. A previous study has highlighted a synergetic effect between the electrocatalyst's electronic and geometrical/morphological properties, revealing that oxide morphology significantly impacts ORR selectivity and $H_2O_2$ electrosynthesis [24].

In our previous work, we thoroughly investigated the synthesis, characterization and 2-electron oxygen reduction reaction (ORR) pathway using nanoflower-like $WO_3$ oxide nanostructures to modify Vulcan XC-72. [28] Our findings demonstrated its great potential, but its application in electro-Fenton processes has yet to be explored. To address this issue and fill the gap in application studies, we report a GDE based on nanoflower-like $WO_3$ oxide-modified Vulcan XC72 carbon for the electrochemical advanced oxidation of ciprofloxacin (CIP). The GDE demonstrated efficient activity for CIP removal, and the degradation mechanism was investigated using high-performance liquid chromatography-mass spectrometry (HPLC-MS) and theoretical calculations. Besides, toxicity was evaluated using the theoretical-based method ECOSAR.

## 2. Methodology

### 2.1. Materials and Chemicals

Tungsten hexachloride (≥ 98%), polytetrafluoroethylene (60 wt% dispersion in H2O), and ciprofloxacin (≥ 98%) were purchased from Sigma-Aldrich. Ethanol, potassium sulfate (≥ 99%), sulfuric acid (≥ 95%), iron (II) sulfate heptahydrate (≥ 99%), $(NH_4)_6Mo_7O_{24}\cdot 4H_2O$ (≥ 81% $MoO_3$) were provided by Synth. All solutions were prepared with ultrapure water (18.2 MΩ cm at 25 ºC) from a Merck Millipore Milli-Q system.

### 2.2. $WO_3$/Vulcan XC 72 catalyst preparation

The catalyst preparation involved a solvothermal synthesis of $WO_3$ nanoparticles and subsequent wet impregnation onto Vulcan XC-72 amorphous carbon (3% w/w oxide: carbon). The detailed synthesis procedure was carried out as described in a previous work [28]. Briefly, for the solvothermal synthesis of $WO_3$, 50 mmol $L^{-1}$ $WCl_6$ in ethanol was transferred to an autoclave with a Teflon inner chamber and heated at 160 ºC for 12 h. The resulting product was washed with water and ethanol, then calcined at 500°C for 1h. Subsequently, an appropriate amount of the synthesized $WO_3$ was suspended in Vulcan carbon via wet impregnation, stirred for 6h, and dried to obtain the 3 % $WO_3/C$ electrocatalyst.

### 2.3. Electrochemical setup

*In situ* $H_2O_2$ electrosynthesis was achieved using a gas diffusion electrode (GDE) cathode prepared by a hot-pressing procedure using 3% $WO_3$/Vulcan XC-72 or pure Vulcan XC-72 containing 20 % (w/w) PTFE dispersion, as previously described by our research group[24,29,30]. Electrogeneration experiments were carried out in an undivided 350 mL electrochemical cell using an Ag|AgCl reference electrode, a Pt or BDD counter electrode (3 $cm^2$), and the fabricated GDE, with continuous $O_2$ supply at 0.2 bar. The electrolyte consisted of 0.1 mol $L^{-1}$ $K_2SO_4$ (pH 3 adjusted with $H_2SO_4$). The electrolysis process was performed by applying 50, 75, and 100 mA $cm^{-2}$ for 120 min. Ciprofloxacin (CIP) degradation experiments were conducted using the same electrochemical cell setup for $H_2O_2$ accumulation. The degradation processes included photolysis, anodic oxidation (AO), AO with electrogenerated $H_2O_2$ (AO-e-$H_2O_2$), electro-Fenton (EF), and photoelectro-Fenton (PEF). The EF and PEF experiments were performed with the addition of 0.50 mM $Fe^{2+}$. For PEF, a mercury UV lamp of $\lambda_{max}$ = 254 nm (Pen-Ray Lamps/UVP 90-0012-01) was immersed in the solution. The initial CIP concentration was 20 mg $L^{-1}$, equivalent to 12 mg $L^{-1}$ total organic carbon (TOC).

### 2.4. Physicochemical characterization

The X-ray diffraction (XRD) was obtained using a D8 Focus diffractometer (Bruker AXS) operating at 40 kV and 40 mA with CuKa radiation in the 20°– 80° 2θ range at a scan rate of 2° $min^{-1}$. The morphologies were investigated using a FESEM JEOL JSM-7401 operating at 5 kV.

### 2.5. Analytical methods

The $H_2O_2$ accumulation was quantified colorimetrically by using $(NH_4)_6Mo_7O_{24}\cdot 4H_2O$ ($2.4 \times 10^{-3}$ mol $L^{-1}$) in $H_2SO_4$ (0.5 mol $L^{-1}$) in a spectrophotometer Varian Cary 50. Fe (II) was quantified by 1,10-phenanthroline colorimetry (Varian Cary 50 spectrophotometer), whereas Total Fe was determined after ascorbic acid reduction of Fe (III), with both species analyzed by the same method. The degradation products of CIP were analyzed by liquid chromatography-tandem mass spectrometry using an LC-20A liquid chromatograph (Shimadzu Co., Japan) and an LC-MS-8030 triple quadrupole mass spectrometer (Shimadzu Co., Japan). The separation was performed on a Shim-pack ODS II column (3 mm i.d. × 100 mm, 2.2 μm) at a temperature of 30 ˚C. The mobile phase consisted of water with 0.1% formic acid (A) and acetonitrile (B). Elution was carried out at a flow rate of 0.6 mL $min^{-1}$ with a gradient program from 10 to 37% (B) within 5 min, followed by 37 to 100% (B) for 3 min, then 100 to 10% (B) for 4 min. The electrospray interface (ESI) was operated in positive ionization mode at 400˚C and 4.0 kV. Full-scan spectrum were obtained in positive-ion mode over the m/z range 100-500. Nitrogen is used as a nebulizer gas at a flow rate of 3 and 15 L $min^{-1}$, respectively.

Total organic carbon (TOC) measurements monitored the mineralization rate. The mineralization process was ceased with sodium sulfite, which was filtered through 0.45 μm PTFE filters from Analytical and analyzed in a TOC-V CPN SHIMADZU analyzer. The TOC analysis was made in triplicate.

### 2.6. Computational simulations

#### 2.6.1. Calculation of Fukui Index

Initially, the theoretical aspect of this study involved calculating the Fukui indices, as these values could be related to possible radical attacks and the generation of new species (degradation products). The three-dimensional structure of the molecule under investigation (ciprofloxacin) was optimized in the Gaussian 09 program[31] , using the CAM-B3LYP density functional, with the 6-31++G(3d,3p) basis set, NPA (Natural Population Analysis) type charges; the medium was simulated with the CPCM (water) implicit solvent model [32–35]. The NPA charges were obtained following the molecular geometry optimization, and the Fukui indices were subsequently calculated [36,37]. These values estimate a molecule's possible reactivity regions (pollutant, drug, etc.). Among the values obtained from the Fukui indices, the f0 index is related to radical attack; the higher this value (more positive), the more susceptible the atom is to radical attack. After analyzing the results of the geometric optimization of the molecule, we carried out the single-point calculations to obtain the energy values for the [1]. Despite the healthcare benefits of antibiotics, the ove

### 2.6.2. Analysis via ECOSAR

After obtaining the Fukui indices, the toxicity of CIP and its byproducts was assessed. For this, the ECOSAR tool was employed to predict the aquatic toxicity of the studied compounds in fish, crustaceans, and green algae. This tool uses structure-activity relationship (SAR) descriptors in linear regression based on LC50 values obtained from tests conducted over 96 hours for fish and 48 hours for daphnids. $LC_{50}$ indicates the concentration required to kill 50% of a sample in an aquatic environment. For green algae, the $EC_{50}$ value is estimated at 96 hours. In this way, CIP and its byproducts (observed from mass spectrometry) were analyzed using the ECOSAR program [38–40].

## 3. Results and discussion.

### 3.1. Structural characterization

The XRD patterns of the synthesized $WO_3$ (Fig. 1(a)) display diffraction peaks that are in excellent agreement with those characteristics of the monoclinic $WO_3$ phase, as indexed by ICSD card No. 17003, confirming the successful formation of a well-crystallized structure without detectable secondary phases. This indicates that the synthesis method employed effectively produced phase-pure $WO_3$ with high crystallinity. In addition, the SEM image (Fig. 1(b)) reveals well-defined, flower-like nanostructures composed of interconnected nanoflakes, organized in a hierarchical architecture. Such morphology is typically associated with a high specific surface area and enhanced accessibility of active sites[41,42], features that can significantly improve the material's electrochemical and catalytic performance.

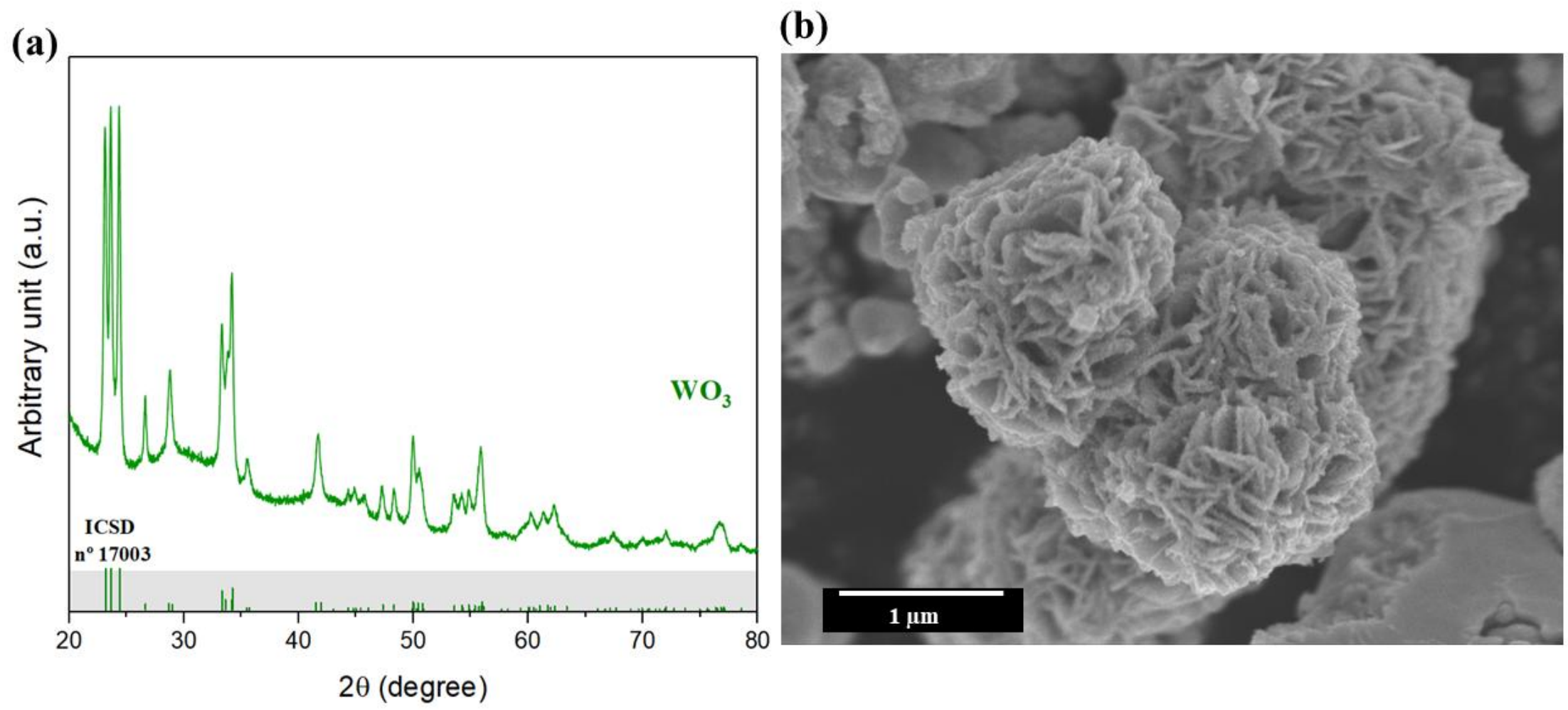

**Figure 1.** (a) XRD pattern and (b) SEM image of the synthesized $WO_3$ nanostructures**.**

### 3.2. Hydrogen Peroxide electrosynthesis evaluation

Electrolytic essays were conducted at different current densities to evaluate the $H_2O_2$ yield performance of the gas diffusion electrodes based on the 3% $WO_3$/C electrocatalyst and bare carbon XC-72. Figure 2(a) illustrates the accumulation of $H_2O_2$ over 120 minutes of electrolysis. The $H_2O_2$ yield increased with the applied current for both GDEs, with $WO_3$/C showing a significant improvement compared to bare Vulcan carbon. Specifically, the 3% $WO_3$/C achieved $H_2O_2$ concentrations of 423 mg $L^{-1}$ and 586 mg $L^{-1}$ at current densities of 50 and 75 mA $cm^{-2}$, respectively. This represents approximately 3.4- and 2.2-fold enhancements over Vulcan carbon at current densities of 50 and 75 mA $cm^{-2}$, respectively. Moreover, this GDE electrocatalyst demonstrated high efficiency at near-industrial current densities; at 100 mA $cm^{-2}$, it accumulated 916 mg $L^{-1}$ of $H_2O_2$, a 1.4-fold improvement over pure Vulcan carbon GDE under the same conditions. In addition to excellent $H_2O_2$ accumulation, the current efficiency of the 3% WO3/C GDE was around 70% at all applied current densities, demonstrating superior performance compared to bare Vulcan GDEs.

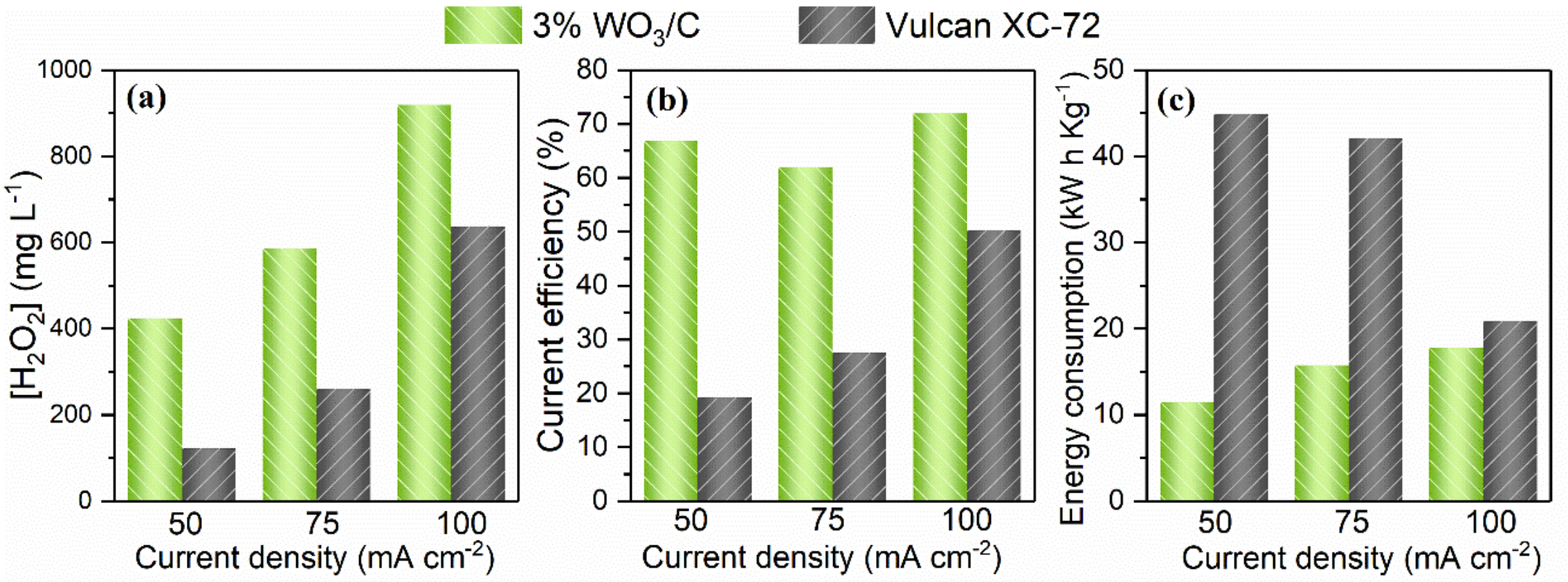

**Figure 2.** (a) $H_2O_2$ yield, (b) current efficiency, and (c) energy consumption achieved with 3% $WO_3$/C GDEs at different applied density currents using 0.1 mol $L^{-1}$ $K_2SO_4$ pH 3 as supporting electrolyte.

The $WO_3$/C GDEs also reduced energy consumption across the tested current range. This improved performance can be attributed to the synergistic effect between Vulcan carbon and $WO_3$ oxide. In particular, the increased number of active sites arises from the oxygen-rich modified surface of Vulcan carbon and the presence of $WO_3$ oxide, which was concluded in a previous study [28].

**3.2. Ciprofloxacin degradation performance**

The EF process in CIP degradation evaluated the catalytic performance of the WO3/C GDE. Several oxidation processes were investigated to better understand activity, as shown in Fig. 3(a). The photolysis, anodic oxidation (AO), and anodic oxidation with $H_2O_2$ electrogenerated media (AO-e-$H_2O_2$) degraded 33.1%, 40.2%, and 43.7%, respectively. On the other hand, CIP degradation under EF-based processes shows two behaviors: the first stage (0-30 minutes) shows a faster degradation rate for both processes, EF and UV-PEF. In the first 30 minutes, the ●OH radical rapidly formed via the Fenton reaction, driving the fast kinetics for the EF processes, displaying a degradation of about 70%. The EF process is highlighted as rapidly degrading. Nevertheless, after 30 minutes, the EF process has poor activity in the second stage (from 30 min to 90 min). In contrast, the UV lamp-assisted process maintained its degradation activity, achieving complete degradation of the CIP solution.

This two-step performance was evidenced by a kinetic-fitted profile; the pseudo-first-order rate constants are shown in Fig. 3(b) and (c), respectively. In the first 30 min of reaction, EF and PEF displayed a kinetic rate constant of 0.038 and 0.045 $min^{-1}$, respectively. At the second stage reaction, the EF rate decreased constantly 10-fold to around (0.003 $min^{-1}$). Meanwhile, the PEF maintained a high kinetic rate reaction (0.039 $min^{-1}$). At the first stage, steady *in-situ* $H_2O_2$ electrogeneration (reaction 1) and the initial $Fe^{2+}$ availability (reaction 2) governs the CIP removal process in both processes.

However, $Fe^{2+}$ is gradually oxidized to Fe3+ via the Fenton reaction. Due to limited cathodic recovery (reaction 3), the Fenton reaction is limited by the availability of free $Fe^{2+}$ species. The generation and accumulation of CIP byproducts initiate parallel and competing oxidation processes that can also hinder the kinetics of the response [43]. [44–47]

$$O_2 + 2e^- + 2H^+ \rightarrow H_2O_2 \quad (1)$$

$$H_2O_2 + Fe^{2+} \rightarrow \cdot OH + Fe^{3+} + OH^- \quad (2)$$

$$Fe^{3+} + e^- \rightarrow Fe^{2+} \quad (3)$$

$$H_2O_2 + hv \rightarrow 2\ \cdot OH \quad (4)$$

$$Fe(OH)^{2+} + hv \rightarrow Fe^{2+} + \cdot OH \quad (5)$$

$$Fe(OOCR)^{2+} + hv \rightarrow Fe^{2+} + CO_2 + R\cdot \quad (6)$$

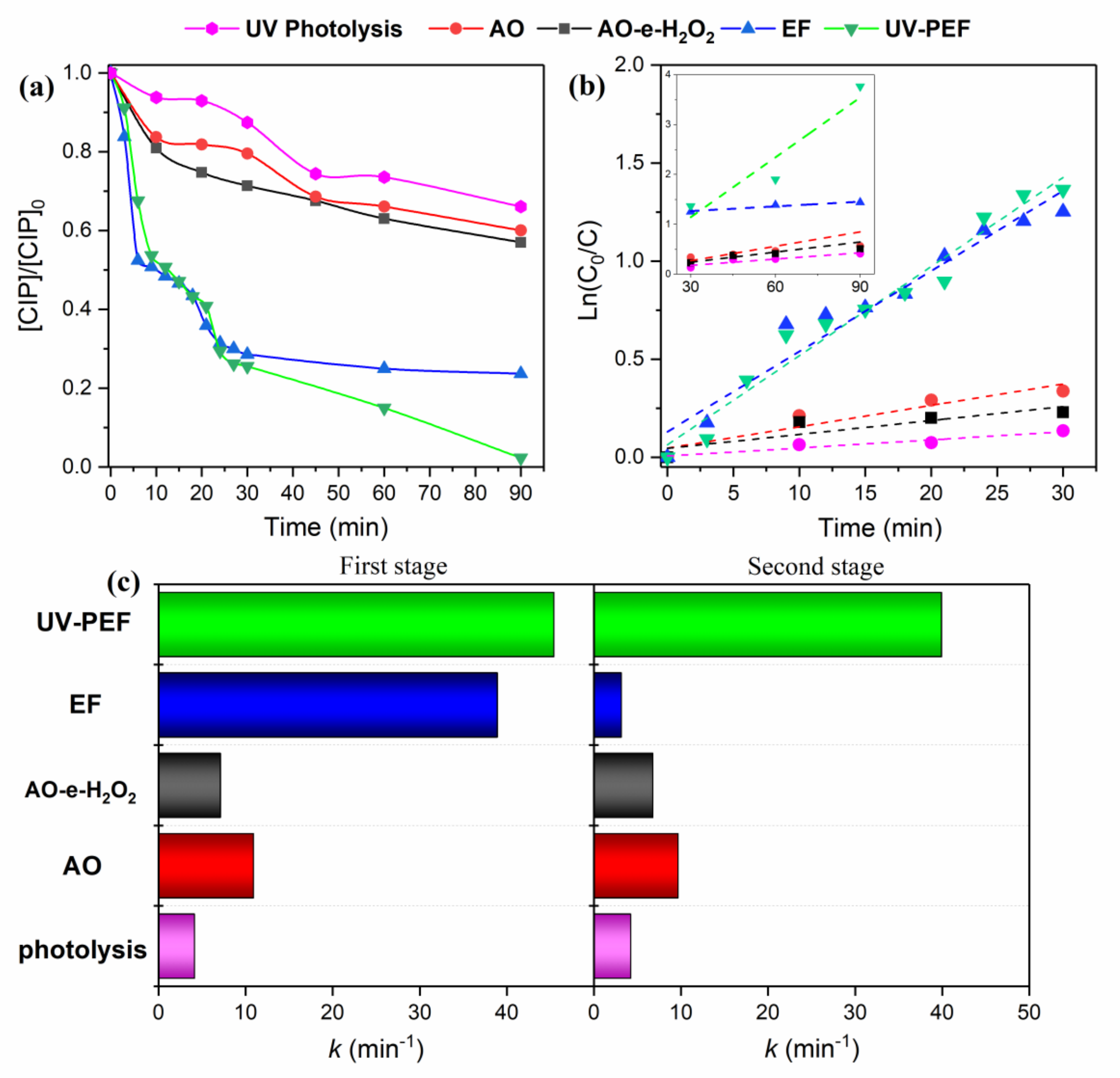

**Figure 3.** (a) CIP degradation, (b) kinetic analysis assuming a pseudo-first-order reaction (inset 2nd stage fitting), (c) kinetic constants using with 3% $WO_3$/C GDEs and Pt anode under different processes using 0.1 mol $L^{-1}$ $K_2SO_4$ (pH 3 adjusted with $H_2SO_4$) as supporting electrolyte.

In this context, the evolution of iron species during the PEF and EF processes was systematically monitored, as presented in Fig. 4(a). These results demonstrate that UV irradiation modestly but noticeably improves $Fe^{2+}$ regeneration compared with the standalone EF process. Moreover, the PEF system exhibited a reduced residual $H_2O_2$, as shown in Fig. 4(b). This behavior suggests that UV irradiation (UV-PEF process) provides an additional synergistic pathway for hydroxyl radical generation, primarily through the photolysis of $H_2O_2$ (Eq. 4). UV light can also promote the dissociation of ferric complexes (Eqs. 5 and 6) [44,46], thereby regenerating free $Fe^{2+}$ ions and enhancing subsequent •OH formation. Collectively, these effects contribute to a more efficient oxidative environment and improved overall system performance

TOC mineralization rates were also utilized to assess the feasibility of potential applications. Based on previous results, CIP mineralization rates were investigated during the PEF process using two distinct anodes: Pt and BDD. The BDD electrode is regarded as one of the most efficient materials for the sustained electrogeneration of •OH in electrochemical water treatment processes [48]. From Fig. 4(c), the TOC removal rate reached 38.5% and 66.0% within 90 min for the PEF-Pt and PEF-BDD, respectively. As discussed, the Pt anode resulted in low CIP degradation, achieving only 9% of CIP mineralization, indicating a limited TOC removal by Pt anodic oxidation. This suggests that the TOC removal in the PEF-Pt process is primarily due to the electro-Fenton reaction. Antonin *et al.* reported similar results when comparing Pt and BDD anodes for CIP degradation [11]. The superior oxidation capability of the BDD material was evident when coupled with the $WO_3$/C GDE, achieving synergistic TOC removal through combined anodic oxidation and photoelectron-Fenton process in just 90 min

[49–56]

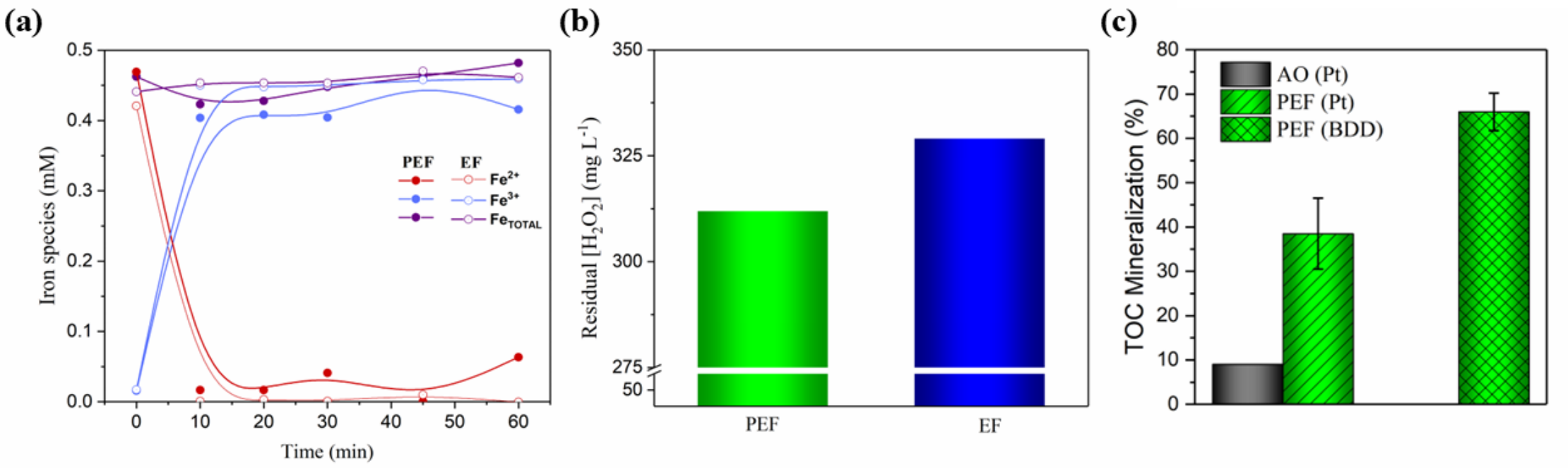

**Figure 4.** (a) Dissolved Fe species evolution, and (b) residual $H_2O_2$ over PEF and EF processes, and (c) TOC mineralization achieved using 3% $WO_3$/C GDEs and Pt or BDD anode under different processes using 0.1 mol $L^{-1}$ $K_2SO_4$ (pH 3 adjusted with $H_2SO_4$) as supporting electrolyte

### 3.3. Experimental proposed ciprofloxacin degradation pathway

The degradation processes applied in this study promoted two degradation pathways for the ciprofloxacin molecule, as shown in Fig 5. Table 1 summarizes the identified intermediates and mass spectra for these products in the supporting material. One of the proposed pathways occurred via •OH radical attack on the piperazine ring, associated or not with the defluorination process, and an attack on the quinolone ring characterized the second. Defluorination of the CIP molecule by hydroxyl radical attack is facilitated due to the high electron density of fluorine, and cleavage of the piperazine ring (Pathway 1) has often been described in the literature [57,58] by breaking the C-C bond, proximal to the quaternary nitrogen, or the C-N bonding, in the aromatic amine heterostructure.

The formation of the degradation product DP 1 of *m/z* 413 is proposed to be due to the cleavage of the C-N bond of the piperazine ring together with the hydroxylation and oxidation of the quinolone and piperazine rings [59]. Sequential dehydroxylation of the quinolone ring of DP1 generated DP2 (*m/z* 391) and DP3 (*m/z* 275) [60,61]. In DP2, there is also oxidation of the piperazine ring and cyclopropane with cleavage of the latter. In forming DP3, the piperazine ring undergoes continued oxidation, losing two $C_2H_6O$ mobility. The removal of the $NO_2$ group and cleavage of the cyclopropane ring from DP3 yields DP4 with *m/z* 218. Decarboxylation and removal of the $C_2H_5$ group from the cyclopropane ring in DP4 are among the possible proposals for the formation of DP5 (*m/z* 146) [62]. DP6 with *m/z* 208 is generated by breaking the C-C bond of the piperazine ring together with the decarboxylation steps and loss of the cyclopropane ring [63].

The attack of hydroxyl radicals on the piperazine ring, without the defluorination step, is the proposed formation of DP 7 (*m/z* 279), DP 8 (*m/z* 224), and DP 9 (*m/z* 222) [64–66]. These compounds are generated by the partial fragmentation or total loss of the piperazine and/or cyclopropane rings. Hydroxylation of the carbon atom adjacent to the cyclopropyl group, reported as one of the most reactive sites for hydroxyl group addition in ciprofloxacin, is also observed in DP 7 and DP 8 [61,65,67].

The degradation pathway via the quinolone ring (Pathway 2) occurred initially by the attack of the oxidizing radical on the C=C bond of the quinolone ring, promoting hydroxylation, oxidation, and opening of the respective ring, and cleavage of the cyclopropane ring, forming the degradation product DP10 (*m/z* = 354) [64]. Subsequent oxidation of the quinolone ring and removal of the $C_2H_5$ group from the cyclopropane ring and $C_4H_{11}N$ from the piperazine ring gave rise to DP11 with *m/z* 259 [68,69].

The structure of DP12 (*m/z* 183) shows that both cited routes occurred simultaneously by the cleavage of the piperazine ring via the breaking of the C-C bond and the degradation of the quinolone group. The attack of OH on the quinolone dual-ring structure generated the defluorination, hydroxylation, and fragmentation of the first ring and the decarboxylation, along with the loss of the carbonyl and cyclopropane groups in the amine ring [70].

Table 1. LC-ESI-MS/MS data on ciprofloxacin degradation products.

| **Compound** | **tR (min)** | **$[M+H]^+$** | **LC-MS/MS fragments *(m/z)*** |
|---|---|---|---|
| DP 1 | 8.7 | 413 | 383, 145, and 105 |
| DP 2 | 1.0 | 391 | 279, 235, 143, and 101 |
| DP 3 | 0.7 | 275 | 191, 123, and 105 |
| DP 4 | 3.4 | 218 | 142, 105, and 101 |
| DP 5 | 7.9 | 146 | 105, and 100 |
| DP 6 | 0.6 | 208 | 167, 122, and 102 |
| DP 7 | 7.3 | 279 | 146, 105, and 101 |
| DP 8 | 0.3 | 224 | 183, 130, 110, and 89 |
| DP 9 | 9.7 | 222 | 145, 121, and 105 |
| DP 10 | 8.3 | 354 | 306, and 100 |
| DP 11 | 0.6 | 259 | 243, 191, 146, and 121 |
| DP 12 | 0.5 | 183 | 167, 110, and 102 |

**Figure 5.** Main reaction pathway proposed for the degradation of ciprofloxacin by electrochemical process.

### 3.4. Simulation of the degradation pathways and toxicity assessment

Table 2 displays the calculated Fukui indices for ciprofloxacin, and Figure 6 illustrates the central atoms of CIP involved in radical attack.

**Table 2**. Charges of the species N, N+1, N-1 and the Fukui index ($f^0$) for ciprofloxacin in a solvated medium (water)

| Atom | qiN | qiN +1 | qiN -1 | f0 |
|---|---|---|---|---|
| O1 | -0.73 | -0.79 | -0.80 | -0.01 |
| C2 | 0.83 | 0.99 | 0.98 | 0.00 |
| O3 | -0.62 | -0.71 | -0.74 | -0.01 |
| C4 | -0.35 | -0.44 | -0.31 | 0.07 |
| C5 | 0.12 | 0.25 | 0.04 | -0.11 |
| N6 | -0.39 | -0.49 | -0.56 | -0.03 |
| C7 | -0.10 | -0.07 | -0.04 | 0.01 |
| C8 | -0.46 | -0.42 | -0.43 | 0.00 |
| C9 | -0.47 | -0.43 | -0.44 | 0.00 |
| C10 | 0.21 | 0.23 | 0.23 | 0.00 |
| C11 | -0.29 | -0.16 | -0.32 | -0.08 |
| C12 | 0.15 | 0.11 | 0.14 | 0.01 |
| N13 | -0.49 | -0.08 | -0.61 | -0.27 |
| C14 | -0.27 | -0.26 | -0.20 | 0.03 |
| C15 | -0.29 | -0.22 | -0.22 | 0.00 |
| N16 | -0.72 | -0.73 | -0.76 | -0.01 |
| C17 | -0.29 | -0.22 | -0.22 | 0.00 |
| C18 | -0.27 | -0.25 | -0.19 | 0.03 |
| C19 | 0.38 | 0.48 | 0.40 | -0.04 |
| F20 | -0.35 | -0.38 | -0.41 | -0.02 |
| C21 | -0.25 | -0.22 | -0.26 | -0.02 |
| C22 | -0.18 | -0.09 | -0.19 | -0.05 |
| C23 | 0.52 | 0.66 | 0.37 | -0.14 |
| O24 | -0.68 | -0.71 | -0.93 | -0.11 |
| H25 | 0.54 | 0.54 | 0.54 | 0.00 |
| H26 | 0.27 | 0.26 | 0.22 | -0.02 |
| H27 | 0.27 | 0.26 | 0.23 | -0.01 |
| H28 | 0.26 | 0.25 | 0.23 | -0.01 |
| H29 | 0.26 | 0.24 | 0.23 | 0.00 |
| H30 | 0.26 | 0.24 | 0.23 | 0.00 |
| H31 | 0.26 | 0.24 | 0.23 | -0.01 |
| H32 | 0.26 | 0.26 | 0.23 | -0.01 |
| H33 | 0.24 | 0.27 | 0.21 | -0.03 |
| H34 | 0.25 | 0.26 | 0.22 | -0.02 |
| H35 | 0.24 | 0.25 | 0.21 | -0.02 |
| H36 | 0.23 | 0.22 | 0.20 | -0.01 |
| H37 | 0.40 | 0.40 | 0.39 | -0.01 |
| H38 | 0.24 | 0.25 | 0.21 | -0.02 |
| H39 | 0.23 | 0.22 | 0.20 | -0.01 |
| H40 | 0.26 | 0.26 | 0.23 | -0.02 |
| H41 | 0.24 | 0.26 | 0.20 | -0.03 |
| H42 | 0.28 | 0.28 | 0.25 | -0.02 |

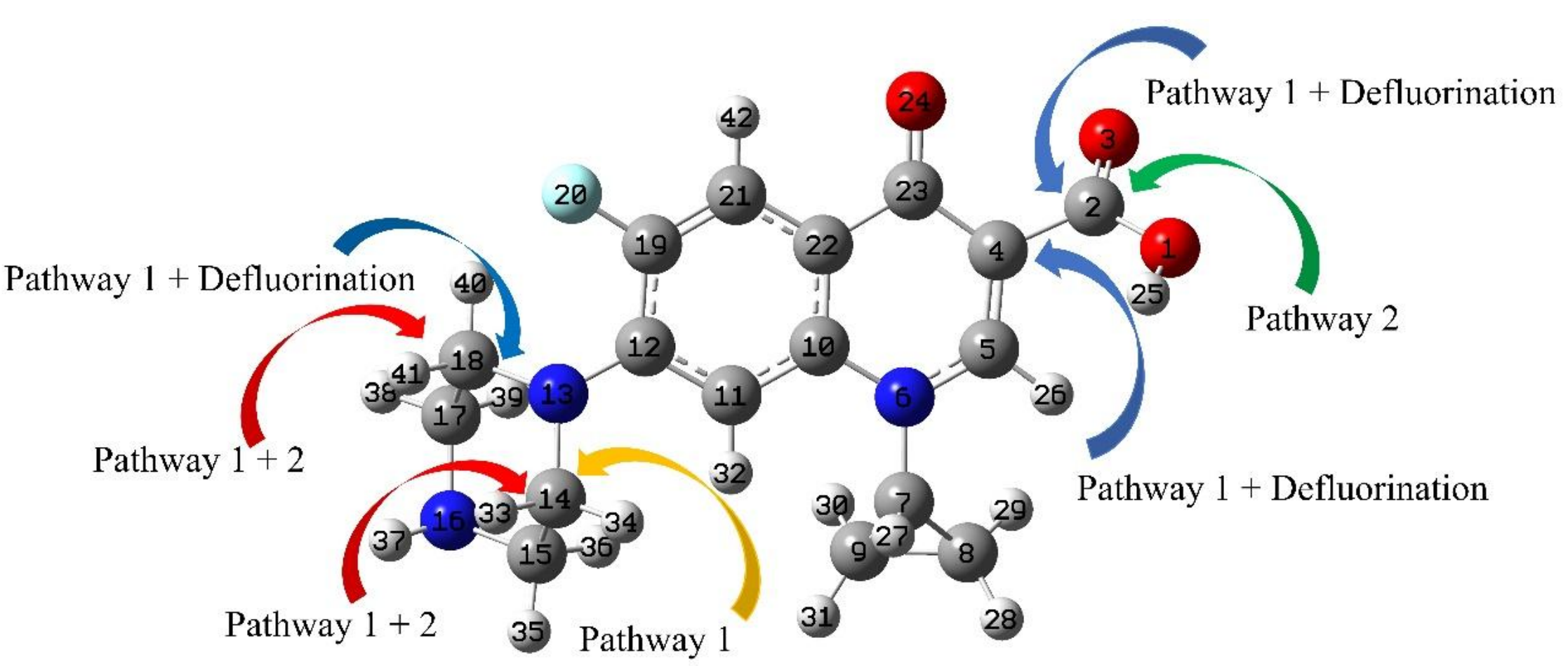

**Figure 6.** The main sites of reactivity involve the hydroxyl radical attacking ciprofloxacin.

Regarding one of the possible byproducts proposed from the electrochemical degradation of ciprofloxacin, the Fukui index (f0) results (Table 2) indicate that the most susceptible atoms to radical attack are C4, C14, and C18. The results presented in Figure 6 identify the most reactive regions, which corroborate the experimental findings.

The radical attack at C4 suggests the possible formation of byproducts via path 2 (DP11) and path 1 + defluorination (DP1), both involving a hydroxyl group at C4. The radical attack suggested by the Fukui index analysis at C14 and C18 indicates the formation of byproducts DP7, DP2, and DP3. After evaluating possible degradation routes, the Fukui index was used to verify the byproducts formed. The results obtained for the other byproducts are presented in the supplementary data.

For DP1, the calculated values of the Fukui index for the atoms C26 and C28 (0.11 and 0.14, respectively) indicate the possible formation of by-product DP2. Regarding the formation of the DP3 by-product, the f0 values for N7 and C11 (0.13 and 0.05, respectively) at DP1 suggest the reactivity of this region, which may break the bonds between N7 and C4, C2, C10, and C12, generating the DP3 by-product.

Considering the formation of DP4, the $f^0$ values obtained for atoms C11 and C14 (0.09 and 0.06, respectively) may indicate a bond break between C12 and N13, forming DP4. DP5 formation occurs via DP4, as evidenced by calculations of the Fukui index. The positive $f^0$ value for atom C4 (0.10) may be related to breaking the bond between C4 and C2. Another broken bond is observed between N6 (0.12) and C7; this rupture can be linked to the positive $f^0$ values for N6 (0.12) and C5 (0.05). Finally, the formation of DP6 may be related to the $f^0$ value for C4 (0.07), since this atom may be more susceptible to radical attack. In this way, the calculations of the Fukui index corroborate the byproducts generated experimentally for Pathway 1 + Defluorination.

For pathway 1, the formation of the DP7 by-product is described above. Regarding the generation of the DP8 by-product, the atoms susceptible to radical attack in the DP7 molecule are corroborated by the experimental results, as the Fukui indices for the C2 and N7 atoms are positive (C2 = 0.08 and N7 = 0.07). These results suggest that the bond between N3 and C4 is broken, forming the byproducts DP8 and DP9. Regarding pathway 2, ciprofloxacin shows a positive $f^0$ value for the C4 atom (0.07), suggesting that this region is reactive to radical attack, where more than one hydroxyl radical can form DP10. Finally, for the formation of DP10, the analysis of Fukui indices indicates that the reactivity regions associated with DP11 formation are located around the N13 and C30 atoms. Due to the positive $f^0$ values in these regions, this analysis suggests that the bond between N13, C14, and C18 may rupture, possibly leading to ring-breaking. For the C30 atom, based on the reactivity of this region at DP10, a hydroxyl is possibly formed from a radical attack, as seen in DP11. This transformation highlights the intricate interplay between bond dynamics and radical interactions within the molecular structure.

Finally, the CIP degradation may form the by-product DP12 from reactivity regions involving the C4, C7, C14, and C18 atoms. The $f^0$ value for C4 (0.07) suggests that the radical attack contributes to breaking the C4-C2 bond. The positive $f^0$ value for C7 (0.01) suggests that this region may be related to the rupture of the N6-C7 bond. Finally, when attacked by a radical, the regions around the atoms C14 and C18 may contribute to breaking the N13-C14 bond, corroborating the experimental results. Therefore, based on the Fukui index analysis, we evaluated the reactivity regions of CIP and the byproducts observed in the experiments.

Overall, the calculated Fukui indices provide valuable insights into the molecular reactivity of ciprofloxacin and its byproducts, aligning with experimental degradation pathways. The results confirm the role of radical attacks at specific atomic sites in promoting the observed transformation processes, reinforcing the mechanistic understanding of ciprofloxacin degradation.

After obtaining the Fukui index and analyzing potential degradation pathways of ciprofloxacin, the ECOSAR tool was used to assess the aquatic toxicity of ciprofloxacin and its byproducts. The ECOSAR values were classified according to the table in the study by Reuschenbach et al. [36], which correlates LC50 and EC50 values from the ECOSAR model with aquatic toxicity levels. According to this parameter, a molecule can be classified as very toxic ($\leq$ 1 $mgL^{-1}$), toxic (> 1 and $\leq$ 10 $mgL^{-1}$), harmful (> 10 and $\leq$ 100 $mgL^{-1}$), and harmless (> 100 $mgL^{-1}$). The acute and chronic toxicity values and the levels of ciprofloxacin and the byproducts are displayed in Fig 7.

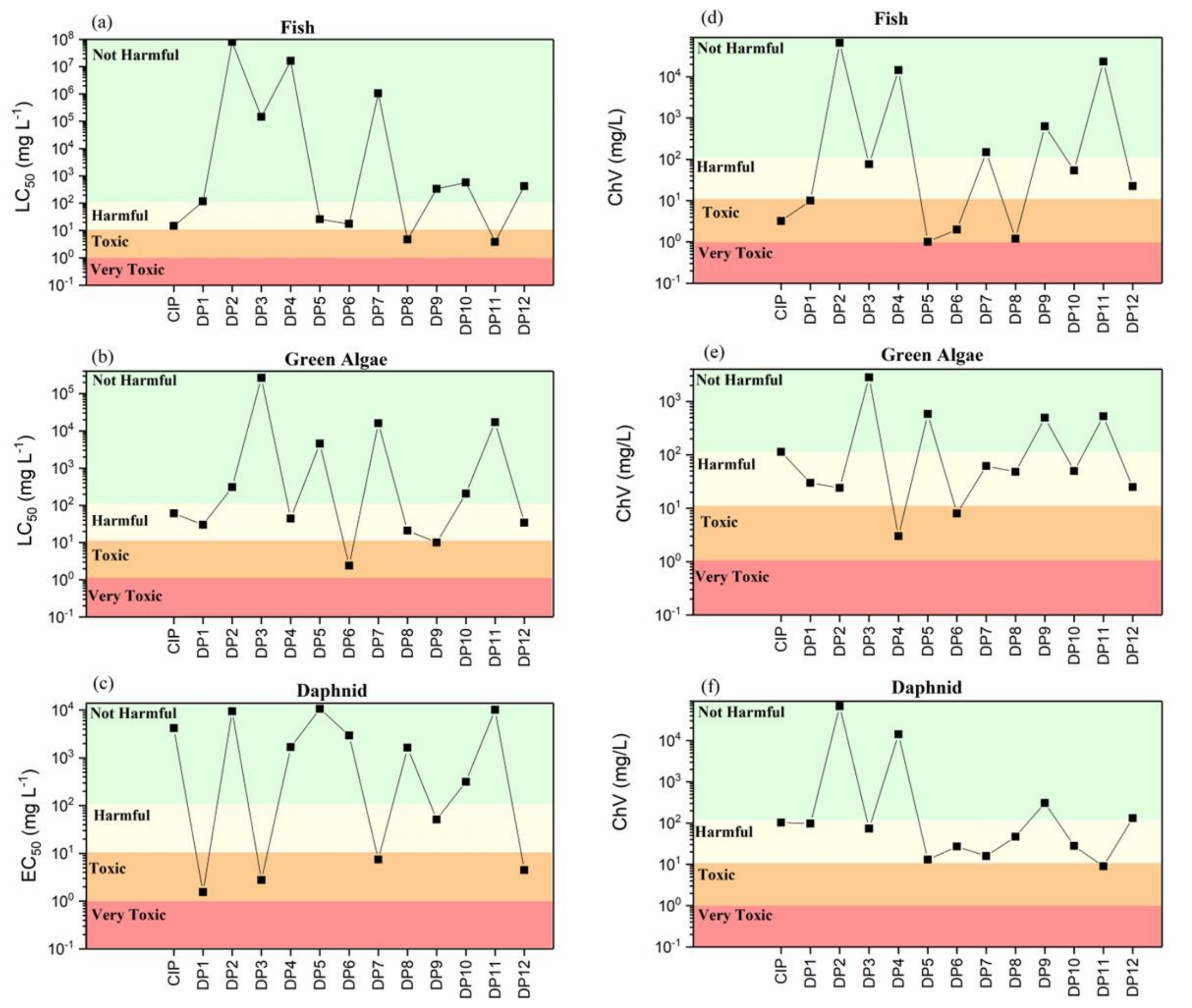

**Figure 7.** The (a-c) acute and (d-f) chronic toxicity of CIP its degradation products predicted by ECOSAR procedure

The values presented in Fig. 7 show that CIP is toxic at chronic levels and harmful at acute levels. For green algae, this substance is harmful at acute levels and harmless at chronic levels. For daphnids, CIP is harmless. The toxicity data for all byproducts obtained from the degradation routes are presented in Table S13. The by-product DP10 is not harmful to the three organisms for acute toxicity, but for chronic toxicity, the values indicate that DP10 is harmful; this by-product shows no toxicity. Another product that is neither very toxic nor toxic is DP9, which, in the acute phase, is not harmful to fish but harmful to green algae and daphnids. For chronic toxicity, DP9 is harmless to all three organisms.

Finally, DP2 is possibly the byproduct with the least effect on aquatic environments, since its acute toxicity values for fish, green algae, and daphnids are harmless. For chronic toxicity, the ECOSAR results indicate that this byproduct is harmful only to green algae. The values for fish and daphnids show no toxicity. The values presented in the Supplementary Material indicate that the other byproducts exhibit varying levels of toxicity, with some toxic byproducts affecting some organisms. Therefore, DP2, DP9, and DP10 do not exhibit high toxicity or toxicity to fish, green algae, or daphnids, suggesting that obtaining these byproducts from CIP degradation might not impact the environment.

## 4. Conclusions

This study demonstrated enhanced electrosynthesis of hydrogen peroxide using 3% WO3/C gas-diffusion electrodes and their application for the degradation of ciprofloxacin solutions. Incorporating 3% $WO_3$ significantly improved the $H_2O_2$ yield compared to bare Vulcan carbon, achieving up to 916 mg $L^{-1}$ at 100 mA $cm^{-2}$, with an efficiency of approximately 70% across various current densities. The synergistic interaction between $WO_3$ and Vulcan carbon increased the number of active sites, enhancing catalytic performance and reducing energy consumption. The results demonstrate that the photoelectron-Fenton process was highly efficient in CIP degradation. UV irradiation significantly improved mineralization rates by enhancing $Fe^{2+}$ regeneration. CIP degradation pathways were proposed based on experimental analyses and theoretical calculations, driven primarily by hydroxyl radical attacks on the piperazine and quinolone rings. Defluorination and subsequent oxidation led to the formation of various intermediate products, which were further validated through Fukui index calculations. These computational analyses confirmed the key reactive sites in the CIP molecule, aligning with the experimentally identified degradation intermediates. Finally, after the electrocatalytic oxidation treatment, the resulting wastewater exhibited reduced environmental impact levels, supported by ECOSAR simulation data. Overall, the results highlight the potential of $WO_3$/C GDEs for sustainable electrochemical $H_2O_2$ production and their applicability in advanced oxidation processes. The findings provide insights into optimizing electro-Fenton treatments for the removal of pharmaceutical pollutants, paving the way for further improvements in wastewater treatment technologies.

**CRediT authorship contribution statement**

**João Paulo C. Moura**: Conceptualization, Investigation, Formal analysis, Data curation, and Writing - original draft; **Erica S. Conrado, Michel O. Almeida**: Software, Formal analysis, and Writing - original draft; **Renata Colombo, Rafael Sotana, Ana M. P. Neto:** : Formal analysis and Writing - review & editing; **Vanessa S. Antonin, Caio M. Fernandes and, Aline B. Trench:** Validation and Writing – review & editing; **Kathia M. Honorio:** Validation and Writing – review & editing; **Mauro C. Santos** Writing - review & editing, supervision and funding acquisition.

**Declaration of Competing Interest**

The authors declare that they have no known competing financial interests or personal relationships that could have appeared to influence the work reported in this paper

**Acknowledgments**

The authors wish to thanks to the financial support of the following Brazilian Research Funding Institutions: Fundação de Amparo à Pesquisa do Estado de São Paulo (FAPESP, 2021/05364-7, 2017/10118-0, 2021/14394-7, 2022/10484-4 and, 2022/12895-1), Coordenação de Aperfeiçoamento de Pessoal de Nível Superior (CAPES) and CNPq (303943/2021-1, 150840/2023-3, 308663/2023–3, 402609/2023–9).

## References

[1] Bhatt S, Chatterjee S. Fluoroquinolone antibiotics: Occurrence, mode of action, resistance, environmental detection, and remediation – A comprehensive review. Environmental Pollution 2022;315. https://doi.org/10.1016/j.envpol.2022.120440.

[2] World Health Organization (WHO). Global Antimicrobial Resistance and Use Surveillance System (GLASS) Report. Geneva: 2021.

[3] World Health Organization. ANTIMICROBIAL RESISTANCE: Global Report on Surveillance. Geneva: 2014.

[4] Cheng DL, Ngo HH, Guo WS, Liu YW, Zhou JL, Chang SW, et al. Bioprocessing for elimination antibiotics and hormones from swine wastewater. Science of the Total Environment 2018;621:1664–82. https://doi.org/10.1016/j.scitotenv.2017.10.059.

[5] Soares PA, Silva TFC V, Arcy AR, Souza SMAGU, Boaventura RAR, Vilar VJP. Assessment of AOPs as a polishing step in the decolourisation of bio-treated textile wastewater: Technical and economic considerations. J Photochem Photobiol A Chem 2016;317:26–38. https://doi.org/10.1016/j.jphotochem.2015.10.017.

[6] Paz EC, Aveiro LR, Pinheiro VS, Souza FM, Lima VB, Silva FL, et al. Evaluation of H2O2 electrogeneration and decolorization of Orange II azo dye using tungsten oxide nanoparticle-modified carbon. Appl Catal B 2018;232:436–45. https://doi.org/10.1016/J.APCATB.2018.03.082.

[7] Paz EC, Pinheiro VS, Aveiro LR, Souza FL, Lanza MR V., Santos MC. Hydrogen Peroxide Electrogeneration by Gas Diffusion Electrode Modified With Tungsten Oxide Nanoparticles for Degradation of Orange II and Sunset Yellow FCF Azo Dyes. J Braz Chem Soc 2019;30:1964–75.

[8] Mohammadi M, Davarnejad R, Sillanpää M. A novel catalyst based on zero-valent iron nanoparticles for assisting electro-fenton process applied to a toxic

wastewater. Results in Engineering 2024;24. https://doi.org/10.1016/j.rineng.2024.102938.

[9] Da IC, Soares C, Oriol R, Ye Z, Martínez-Huitle CA, Cabot PL, et al. Photoelectro-Fenton treatment of pesticide triclopyr at neutral pH using Fe(III)-EDDS under UVA light or sunlight n.d. https://doi.org/10.1007/s11356-020-11421-8/Published.

[10] Pinheiro VS, Paz EC, Aveiro LR, Parreira LS, Souza FM, Camargo PHC, et al. Mineralization of paracetamol using a gas diffusion electrode modified with ceria high aspect ratio nanostructures. Electrochim Acta 2019;295:39–49. https://doi.org/10.1016/j.electacta.2018.10.097.

[11] Antonin VS, Santos MC, Garcia-Segura S, Brillas E. Electrochemical incineration of the antibiotic ciprofloxacin in sulfate medium and synthetic urine matrix. Water Res 2015;83:31–41. https://doi.org/10.1016/j.watres.2015.05.066.

[12] Antonin VS, Aquino JM, Silva BF, Silva AJ, Rocha-Filho RC. Comparative study on the degradation of cephalexin by four electrochemical advanced oxidation processes: Evolution of oxidation intermediates and antimicrobial activity. Chemical Engineering Journal 2019;372:1104–12. https://doi.org/10.1016/j.cej.2019.04.185.

[13] Xia P, Ye Z, Zhao L, Xue Q, Lanzalaco S, He Q, et al. Tailoring single-atom FeN4 moieties as a robust heterogeneous catalyst for high-performance electro-Fenton treatment of organic pollutants. Appl Catal B 2023;322. https://doi.org/10.1016/j.apcatb.2022.122116.

[14] Barndõk H, Blanco L, Hermosilla D, Blanco Á. Heterogeneous photo-Fenton processes using zero valent iron microspheres for the treatment of wastewaters contaminated with 1,4-dioxane. Chemical Engineering Journal 2016;284:112–21. https://doi.org/10.1016/j.cej.2015.08.097.

[15] Trench AB, Fernandes CM, Moura JPC, Lucchetti LEB, Lima TS, Antonin VS, et al. Hydrogen peroxide electrogeneration from O2 electroreduction: A review focusing on carbon electrocatalysts and environmental applications. Chemosphere 2024;352. https://doi.org/10.1016/j.chemosphere.2024.141456.

[16] Gentil TC, Minichova M, Briega-Martos V, Pinheiro VS, Souza FM, Paulo C. Moura J, et al. Stability of supported Pd-based ethanol oxidation reaction electrocatalysts in alkaline media. J Catal 2024;440:115816. https://doi.org/10.1016/J.JCAT.2024.115816.

[17] Valim RB, Carneiro JF, Lourenço JC, Hammer P, dos Santos MC, Rodrigues LA, et al. Synthesis of Nb2O5/C for H2O2 electrogeneration and its application for the degradation of levofloxacin. J Appl Electrochem 2023. https://doi.org/10.1007/s10800-023-01975-z.

[18] Valim RB, Carneiro JF, Lourenço JC, Hammer P, dos Santos MC, Rodrigues LA, et al. Synthesis of Nb2O5/C for H2O2 electrogeneration and its application for

the degradation of levofloxacin. J Appl Electrochem 2023;1:1–15. https://doi.org/10.1007/S10800-023-01975-Z/FIGURES/7.

[19] Valim RB, Trevelin LC, Sperandio DC, Carneiro JF, Santos MC, Rodrigues LA, et al. Using carbon black modified with Nb2O5 and RuO2 for enhancing selectivity toward H2O2 electrogeneration. J Environ Chem Eng 2021;9. https://doi.org/10.1016/j.jece.2021.106787.

[20] Pinheiro VS, Paz EC, Aveiro LR, Parreira LS, Souza FM, Camargo PHC, et al. Ceria high aspect ratio nanostructures supported on carbon for hydrogen peroxide electrogeneration. Electrochim Acta 2018;259:865–72. https://doi.org/10.1016/J.ELECTACTA.2017.11.010.

[21] Antonin VS, Lucchetti LEB, Souza FM, Pinheiro VS, Moura JPC, Trench AB, et al. Sodium niobate microcubes decorated with ceria nanorods for hydrogen peroxide electrogeneration: An experimental and theoretical study. J Alloys Compd 2023;965:171363. https://doi.org/10.1016/j.jallcom.2023.171363.

[22] Machado Fernandes C, Moura JPC, Trench AB, Sotana R, Neto AMP, Santos WG, et al. Electron paramagnetic resonance study of radical species on $NaNbO_3$@$CeO_2$-modified carbon Vulcan XC72 gas diffusion electrode for electrochemical degradation of paracetamol via electro-Fenton. Colloids Surf A Physicochem Eng Asp 2026;728:138712. https://doi.org/10.1016/J.COLSURFA.2025.138712.

[23] Aveiro LR, da Silva AGM, Candido EG, Paz EC, Pinheiro VS, Parreira LS, et al. MnO2/Vulcan-Based Gas Diffusion Electrode for Mineralization of Diazo Dye in Simulated Effluent. Electrocatalysis 2020;11:268–74. https://doi.org/10.1007/s12678-020-00583-1.

[24] Moura JPC, Antonin VS, Trench AB, Santos MC. Hydrogen peroxide electrosynthesis: A comparative study employing Vulcan carbon modification by different MnO2 nanostructures. Electrochim Acta 2023;463. https://doi.org/10.1016/j.electacta.2023.142852.

[25] Aveiro LR, da Silva AGM, Antonin VS, Candido EG, Parreira LS, Geonmonond RS, et al. Carbon-supported MnO2 nanoflowers: Introducing oxygen vacancies for optimized volcano-type electrocatalytic activities towards H2O2 generation. Electrochim Acta 2018;268:101–10. https://doi.org/10.1016/j.electacta.2018.02.077.

[26] Li L, Li J, Fang F, Zhang Y, Zhou T, Zhou C, et al. Efficient H2O2 production from urine treatment based on a self-biased WO3/TiO2-Si PVC photoanode and a WO3/CMK-3 cathode. Appl Catal B 2023;333. https://doi.org/10.1016/j.apcatb.2023.122776.

[27] Bai X, Li Y, Xie L, Liu X, Zhan S, Hu W. A novel Fe-free photo-electro-Fenton-like system for enhanced ciprofloxacin degradation: Bifunctional Z-scheme WO3/g-C3N4. Environ Sci Nano 2019;6:2850–62. https://doi.org/10.1039/c9en00528e.

[28] João Paulo C. Moura, L.E.B. Luchetti, C.M. Fernandes, A.B. Trench, J.M. Almeida, M.C. Santos. Experimental and theoretical studies of WO3/Vulcan XC-72 electrocatalyst enhanced H2O2 yield ORR performed in acid and alkaline medium . Manuscript Submitted for Publication 2024.

[29] Trench AB, Paulo C. Moura J, Antonin VS, Machado Fernandes C, Liu L, Santos MC. Improvement of H2O2 electrogeneration using a Vulcan XC72 carbon-based electrocatalyst modified with Ce-doped Nb2O5. Advanced Powder Technology 2024;35:104404. https://doi.org/10.1016/J.APT.2024.104404.

[30] Antonin VS, Assumpcao MHMT, Silva JCM, Parreira LS, Lanza MRV, Santos MC. Synthesis and characterization of nanostructured electrocatalystsbased on nickel and tin for hydrogen peroxide electrogeneration. Electrochim Acta 2013;109:245–51. https://doi.org/10.1016/j.electacta.2013.07.078.

[31] Sieffert N, Wipff G. Uranyl extraction by N,N-dialkylamide ligands studied using static and dynamic DFT simulations. Dalton Transactions 2015;44:2623–38. https://doi.org/10.1039/C4DT02443E.

[32] Becke AD. Density-functional exchange-energy approximation with correct asymptotic behavior. Phys Rev A (Coll Park) 1988;38:3098. https://doi.org/10.1103/PhysRevA.38.3098.

[33] Ditchfield R, Hehre WJ, Pople JA. Self-Consistent Molecular-Orbital Methods. IX. An Extended Gaussian-Type Basis for Molecular-Orbital Studies of Organic Molecules. J Chem Phys 1971;54:724–8. https://doi.org/10.1063/1.1674902.

[34] Lee C, Yang W, Parr RG. Development of the Colle-Salvetti correlation-energy formula into a functional of the electron density. Phys Rev B 1988;37:785. https://doi.org/10.1103/PhysRevB.37.785.

[35] Cossi M, Rega N, Scalmani G, Barone V. Energies, structures, and electronic properties of molecules in solution with the C-PCM solvation model. J Comput Chem 2003;24:669–81. https://doi.org/10.1002/JCC.10189.

[36] Fukui K, Yonezawa T, Shingu H. A Molecular Orbital Theory of Reactivity in Aromatic Hydrocarbons. J Chem Phys 1952;20:722–5. https://doi.org/10.1063/1.1700523.

[37] Vlahović F, Ognjanović M, Djurdjić S, Kukuruzar A, Antić B, Dojčinović B, et al. Design of an ethidium bromide control circuit supported by deep theoretical insight. Appl Catal B 2023;334:122819. https://doi.org/10.1016/J.APCATB.2023.122819.

[38] Reuschenbach P, Silvani M, Dammann M, Warnecke D, Knacker T. ECOSAR model performance with a large test set of industrial chemicals. Chemosphere 2008;71:1986–95. https://doi.org/10.1016/J.CHEMOSPHERE.2007.12.006.

[39] Mayo-Bean K, Moran K, Meylan B, Ranslow P. ECOSAR Methodology Document 2012.

[40] Khan K, Baderna D, Cappelli C, Toma C, Lombardo A, Roy K, et al. Ecotoxicological QSAR modeling of organic compounds against fish: Application of fragment based descriptors in feature analysis. Aquatic Toxicology 2019;212:162–74. https://doi.org/10.1016/J.AQUATOX.2019.05.011.

[41] Li Y, He G, HuangFu H, Mi Y, Zhang H, Zheng D, et al. Morphology Evolution of NiFe Layered Double-Hydroxide Nanoflower Clusters from Nanosheets: Controllable Structure–Performance Relation for Green Energy Storage. Energy Technology 2024;12:2300749. https://doi.org/10.1002/ente.202300749.

[42] Han SJ, Ameen M, Hanifah MFR, Aqsha A, Bilad MR, Jaafar J, et al. Catalytic Evaluation of Nanoflower Structured Manganese Oxide Electrocatalyst for Oxygen Reduction in Alkaline Media. Catalysts 2020, Vol 10, Page 822 2020;10:822. https://doi.org/10.3390/catal10080822.

[43] Clematis D, Panizza M. Electro-Fenton, solar photoelectro-Fenton and UVA photoelectro-Fenton: Degradation of Erythrosine B dye solution. Chemosphere 2021;270:129480. https://doi.org/10.1016/J.CHEMOSPHERE.2020.129480.

[44] Cornejo OM, Sirés I, Nava JL. Characterization of a flow-through electrochemical reactor for the degradation of ciprofloxacin by photoelectro-Fenton without external oxygen supply. Chemical Engineering Journal 2023;455:140603. https://doi.org/10.1016/j.cej.2022.140603.

[45] Murrieta MF, Brillas E, Nava JL, Sirés I. Solar photoelectro-Fenton-like process with anodically-generated HClO in a flow reactor: Norfloxacin as a pollutant with a particular structure. Sep Purif Technol 2023;308:122893. https://doi.org/10.1016/j.seppur.2022.122893.

[46] Ye Z, Sirés I, Zhang H, Huang YH. Mineralization of pentachlorophenol by ferrioxalate-assisted solar photo-Fenton process at mild pH. Chemosphere 2019;217:475–82. https://doi.org/10.1016/j.chemosphere.2018.10.221.

[47] Antonio da Silva D, Cavalcante RP, Cunha RF, Machulek A, César de Oliveira S. Effective degradation of phenacetin in wastewater by (photo)electro-Fenton processes: Investigation of variables, acute toxicity, and intermediates. J Environ Chem Eng 2024;12:112704. https://doi.org/10.1016/j.chemosphere.2018.05.115.

[48] Jiang Y, Zhao H, Liang J, Yue L, Li T, Luo Y, et al. Anodic oxidation for the degradation of organic pollutants: Anode materials, operating conditions and mechanisms. A mini review. Electrochem Commun 2021;123:106912. https://doi.org/10.1016/J.ELECOM.2020.106912.

[49] Liu Z, Wan J, Ma Y, Wang Y. In situ synthesis of FeOCl@MoS2 on graphite felt as novel electro-Fenton cathode for efficient degradation of antibiotic ciprofloxacin at mild pH. Chemosphere 2021;273. https://doi.org/10.1016/j.chemosphere.2021.129747.

[50] Liu Z jun, Wan J quan, Yan Z cheng, Wang Y, Ma Y wen. Efficient removal of ciprofloxacin by heterogeneous electro-Fenton using natural air–cathode.

Chemical Engineering Journal 2022;433. https://doi.org/10.1016/j.cej.2021.133767.

[51] Ren S, Wang A, Zhang Y, Song Y, Wen Z, Zhang Z. Mechanism insights into H2O2 in situ generation and activation for ciprofloxacin degradation by electro-peroxidation using the ACF@RuO2/Ti cathode. Sep Purif Technol 2024;345. https://doi.org/10.1016/j.seppur.2024.127437.

[52] Wang S, Hu W, Li Y, Zhang D, Hu W, Li Y, et al. Efficient generation of singlet oxygen on Fe2O3/MoO3 Z-type heterojunction for removal of ciprofloxacin from water via photo-electro-Fenton-like system. Chemical Engineering Journal 2024;497. https://doi.org/10.1016/j.cej.2024.154400.

[53] Li X, Yang J, Shi X, Sun Z. N, P co-doped graphite felt cathode for efficient removal of ciprofloxacin in an ascorbic acid-coupled electro-Fenton process: Simultaneously enhancing H2O2 generation and Fe3+/Fe2+ cycling. Environ Res 2025;266. https://doi.org/10.1016/j.envres.2024.120577.

[54] Ahmad A, Kumar P, Kanwar B, Singh SP. Microbial fuel cell powered Fenton degradation of ciprofloxacin from wastewater employing MIL-88B(Fe)-MOF-laser-induced graphene cathode. Environ Res 2026:123807. https://doi.org/10.1016/j.envres.2026.123807.

[55] Fan S, Hou Y, Pan J, Zhu T, Zhang S, Liang T, et al. Cu-Doped V-Based MOF Derivative VO2@Cu-VMOF as a Cathodic Catalyst for Electro-Fenton Degradation of Antibiotics. Small 2025;21. https://doi.org/10.1002/smll.202406523.

[56] Ma B, Li J, Yang C, Wang D. Comparative study of electro-Fenton and photoelectro-Fenton processes using a novel photocatalytic fuel cell electro-Fenton system with g-C3N4@N-TiO2 and Ag/CNT@CF as electrodes. Water Environment Research 2024;96. https://doi.org/10.1002/wer.10946.

[57] Lai C, Zhang M, Li B, Huang D, Zeng G, Qin L, et al. Fabrication of CuS/BiVO4 (0 4 0) binary heterojunction photocatalysts with enhanced photocatalytic activity for Ciprofloxacin degradation and mechanism insight. Chemical Engineering Journal 2019;358:891–902. https://doi.org/10.1016/J.CEJ.2018.10.072.

[58] An T, Yang H, Li G, Song W, Cooper WJ, Nie X. Kinetics and mechanism of advanced oxidation processes (AOPs) in degradation of ciprofloxacin in water. Appl Catal B 2010;94:288–94. https://doi.org/10.1016/J.APCATB.2009.12.002.

[59] Gunawan G, Prasetya NBA, Wijaya RA. Degradation of Ciprofloxacin (CIP) Antibiotic Waste using The Advanced Oxidation Process (AOP) Method with Ferrate (VI) from Extreme Base Electrosynthesis. Trends in Sciences 2023;20. https://doi.org/10.48048/tis.2023.6639.

[60] Xu J, Wang Y, Wan J, Wang L. Facile synthesis of carbon-doped CoMn2O4/Mn3O4 composite catalyst to activate peroxymonosulfate for ciprofloxacin degradation. Sep Purif Technol 2022;287. https://doi.org/10.1016/j.seppur.2022.120576.

[61] He R, Xue K, Wang J, Yan Y, Peng Y, Yang T, et al. Nitrogen-deficient g-C3Nx/POMs porous nanosheets with P–N heterojunctions capable of the efficient photocatalytic degradation of ciprofloxacin. Chemosphere 2020;259. https://doi.org/10.1016/j.chemosphere.2020.127465.

[62] Pan L jia, Li J, Li C xing, Tang X da, Yu G wei, Wang Y. Study of ciprofloxacin biodegradation by a Thermus sp. isolated from pharmaceutical sludge. J Hazard Mater 2018;343:59–67. https://doi.org/10.1016/j.jhazmat.2017.09.009.

[63] Gao Y, Cong S, Yu H, Zou D. Investigation on microwave absorbing properties of 3D C@ZnCo2O4 as a highly active heterogenous catalyst and the degradation of ciprofloxacin by activated persulfate process. Sep Purif Technol 2021;262. https://doi.org/10.1016/j.seppur.2021.118330.

[64] Yakamercan E, Aygün A, Simsek H. Antibiotic ciprofloxacin removal from aqueous solutions by electrochemically activated persulfate process: Optimization, degradation pathways, and toxicology assessment. J Environ Sci (China) 2024;143:85–98. https://doi.org/10.1016/j.jes.2023.08.013.

[65] Du F, Lai Z, Tang H, Wang H, Zhao C. Construction and application of BiOCl/Cu-doped Bi2S3 composites for highly efficient photocatalytic degradation of ciprofloxacin. Chemosphere 2022;287. https://doi.org/10.1016/j.chemosphere.2021.132391.

[66] Bisht K, Kumar G, Dutta RK. Amine-Functionalized Crystalline Carbon Nanodots Decorated on Bi2WO6Nanoplates as Solar Photocatalysts for Efficient Degradation of Tetracycline and Ciprofloxacin. Ind Eng Chem Res 2022;61:16946–61. https://doi.org/10.1021/acs.iecr.2c02635.

[67] Li W, Li S, Tang Y, Yang X, Zhang W, Zhang X, et al. Highly efficient activation of peroxymonosulfate by cobalt sulfide hollow nanospheres for fast ciprofloxacin degradation. J Hazard Mater 2020;389. https://doi.org/10.1016/j.jhazmat.2019.121856.

[68] Zhang R, Peng C, Wang Q, Zou X, Zhao Q, Wang J, et al. Supplementary material For Degradation of ciprofloxacin in UV/NH 2 Cl process: kinetics, mechanism, pathways and DBPs formation. 2023.

[69] Raj SNM, Jothi VK, Rajaram A, Suresh P, Murugan K, Natarajan A. Rational design of α-MnO2/HT-GCN nanocomposite for effective photocatalytic degradation of ciprofloxacin and pernicious activity. Environmental Science and Pollution Research 2023;30:90689–707. https://doi.org/10.1007/s11356-023-28636-0.

[70] O. Silva T, A. Goulart L, Sánchez-Montes I, O. S. Santos G, B. Santos R, Colombo R, et al. Using a novel gas diffusion electrode based on PL6 carbon modified with benzophenone for efficient H2O2 electrogeneration and degradation of ciprofloxacin. Chemical Engineering Journal 2023;455. https://doi.org/10.1016/j.cej.2022.140697.